\documentclass[prc,twocolumn,amssymb,amsmath,amsfonts,aps]{revtex4}
\usepackage{amsmath}  
\usepackage{amsfonts} 
\usepackage{amssymb}
\usepackage{graphicx}   

\begin{document}

\title{A Short Foucault Pendulum Free of Ellipsoidal Precession}
\author{Reinhard A. Schumacher}
\author{Brandon Tarbet}
\affiliation{Department of Physics, Carnegie Mellon University, Pittsburgh, PA 15213}
\email{schumacher@cmu.edu} 
\date{\today}

\begin{abstract}
\noindent A quantitative method is presented for stopping the
intrinsic precession of a spherical pendulum due to ellipsoidal
motion.  Removing this unwanted precession renders the Foucault
precession due to the turning of the Earth readily observable.
The method is insensitive to the size and direction of the
perturbative forces leading to ellipsoidal motion.  We demonstrate
that a short (three meter) pendulum can be pushed in a controlled way
to make the Foucault precession dominant.  The method makes
room-height or table-top Foucault pendula more accurate and practical
to build.
\end{abstract}
\maketitle

\section{Introduction}

L\'eon Foucault built his first pendulum to demonstrate the turning of
the Earth in the basement of a building, using a roughly two meter
long fiber~\cite{fou}.  He also soon recognized the problem arising
from the intrinsic precession of a spherical pendulum caused by
unwanted ellipsoidal motion.  Imperfections in the suspension or
initial conditions of the pendulum generally cause this to quickly
grow to the point that the precession due to the Earth's turning is
overwhelmed.  The pendulum can come to precess in either sense
(clockwise or counterclockwise) at almost any rate, or indeed even
cease all precession.  These practical problems are mitigated in
pendula of great length, and so most are constructed to have lengths
of tens of meters, starting with the celebrated 67 m long device built
by Foucault in Paris in 1851.

The precession of an ideal spherical pendulum with no ellipsoidal
motion is caused by the non-inertial nature of the reference frame
tied to the surface of the Earth.  The so-called Coriolis force
advances the plane of the pendulum's motion by an amount
\begin{equation} \label{GrindEQ__1_} 
\Omega _{F} =\Omega _{Earth} \sin \theta _{latitude}  
\end{equation} 
where $\Omega _{Earth} $ is the sidereal rate of rotation of the
Earth, $\theta _{latitude} $ is the latitude of the pendulum measured
from the equator, and $\Omega _{F} $ is the Foucault precession rate;
many textbooks treat this problem.  Near $40^o$ northern latitude, this
amounts to almost $10^o$ of clockwise advance per hour, for an 18 hour
half-rotation of the pendulum, after which the motion repeats.  This
amount of precession is easily masked, however, by the intrinsic
precession of a less than perfect pendulum that develops some
non-planar ellipsoidal motion.

The construction of a room-height or table-top version of a Foucault
pendulum thus presents a technical challenge, first to minimize the
amount of ellipsoidal motion that accrues as the pendulum swings, and
secondly to compensate in some way for the irreducible amount of this
motion that remains.  In this paper we will first discuss the dynamics
of the spherical pendulum that lead to the problem (Section II).  Then
we introduce a method that stops the ellipsoidal motion of the
pendulum from causing precession, and show that this immunity is, to
first order, independent of the minor axis of the ellipse.  The method
hinges on the observation that pushing the pendulum bob away from the
origin after it passes, rather than either pulling it in or
alternately pulling and pushing it, acts in a way to counter the
unwanted intrinsic precession (Section III).  We then present the
design of a pendulum and a driving mechanism to exploit this method
(Section IV), and demonstrate the validity of this approach by
discussing the supporting experimental results (Section V).  Finally,
we contrast our results with earlier published work on Foucault
pendula and summarize how our method and design are new and unique
(Section VI).

\section{The Problem of Intrinsic Precession}

\begin{figure}[ht]
\begin{center}
\includegraphics[width=0.45\textwidth] {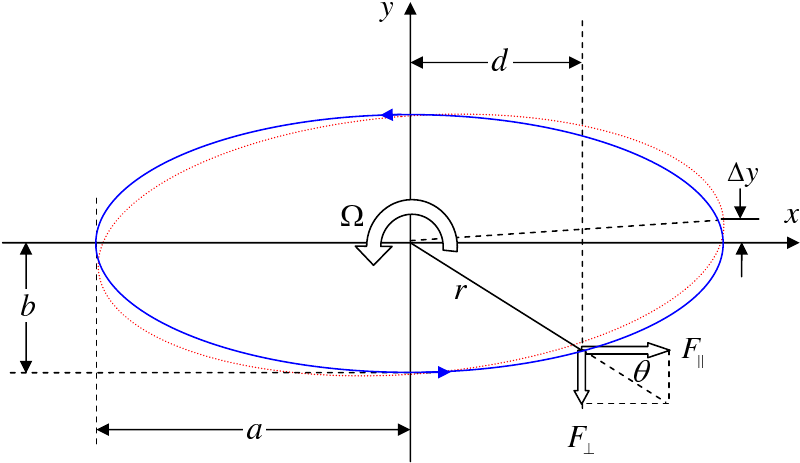}
\caption{
\label{fig:1}
Planar view of the approximate path of a spherical
pendulum with semi-major axis \textit{a} and semi-minor axis
\textit{b} that is moving in a counterclockwise ellipsoid.  The
suspension is centered on the \textit{z} axis above the origin.  The
pendulum is precessing at rate $\Omega$, and in one full cycle the
apex advances by a distance $\Delta y$, as suggested by the light
dotted and rotated ellipse.  The impulsive driving force is applied at
\textit{x }= \textit{d}, and it is resolved into components parallel
and perpendicular to the major axis.  The minor axis can be larger or
smaller, resulting in a \textit{b}-dependent magnitude of the
transverse force $F_{\bot } $ for a fixed longitudinal force
$F_{\parallel } $.  }
\end{center}
\end{figure}

\begin{figure}[ht]
\begin{center}
\includegraphics[scale=0.75]{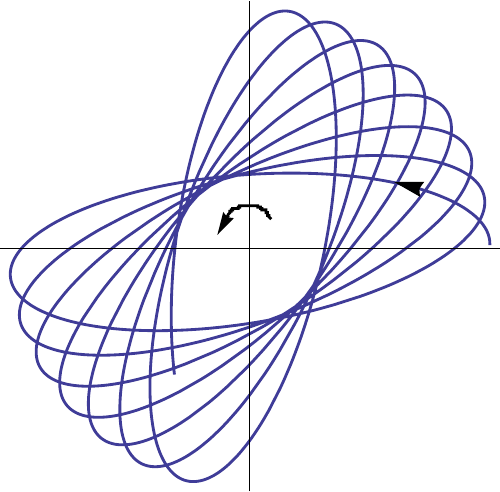}
\caption{
\label{fig:2}
Numerical simulation (using Mathematica) of a spherical
pendulum with a large ratio of semi-minor to semi-major ellipse axes.
The arrows indicate the counterclockwise intrinsic precession, in the
same sense as the motion of the pendulum.}
\end{center}
\end{figure}

The dynamics of the idealized spherical pendulum are determined by the
centrally-directed force of gravity and initial conditions, and lead
to approximately elliptical motion with a semi-major axis \textit{a}
and a semi-minor axis \textit{b}, as shown in Fig.~\ref{fig:1}.  It is
somewhat counter-intuitive but true that the centrally-directed
restoring force of gravity results in a constant intrinsic precession
rate $\Omega$ about the\textit{ z} axis. This precession arises from
the symmetry-breaking condition of having a finite minor axis,\textit{
b}.  The rate is linearly proportional to\textit{ b}, and its sense is
in the same direction as the ellipsoidal motion.  An example of the
exact path of a pendulum with a large ratio of \textit{b/a} is given
in Fig.~\ref{fig:2}, which illustrates a numerical solution of the
equations of motion of a spherical pendulum.

%
%

%
%

Any ellipsoidal motion that develops in a pendulum will result in an
intrinsic precession rate $\Omega$ that is wholly unrelated to the
Foucault precession $\Omega_F$ of Eq \eqref{GrindEQ__1_}.  This has
been worked out in detail by Olsson~\cite{ols1, ols2} and
Pippard~\cite{pip}, among others, and it appears in some textbooks,
for example in Ref~\cite{syn}.  The main result is
\begin{equation} \label{GrindEQ__2_} 
\Omega =\frac{3}{8} \omega _{0} \frac{ab}{L^{2} } ,   
\end{equation} 
where $\omega _{0} =2\pi /T=\sqrt{g/L} $ is the pendulum's angular
frequency, \textit{L} is the length, \textit{T }is the period, and
\textit{g} is the acceleration due to gravity.  The formula is the
lowest-order term in a complicated motion, but it is easily sufficient
for our purpose.  The area of an ellipse, \textit{A}, is given by
$A=\pi ab$, so another way to write the intrinsic precession rate is
\begin{equation} \label{GrindEQ__3_} 
\Omega =\frac{3}{4} \frac{A}{L^{2} T} .  
\end{equation} 
From the first expression, note that the ratio of $\Omega$ (the very
slow intrinsic precession) to $\omega_0$ (the rapid pendular rate) is
proportional to the ratio of the area of the ellipse to the area of a
sphere of radius \textit{L}, that is,
\begin{equation} \label{GrindEQ__4_} 
\frac{\Omega }{\omega _{0} } =\frac{3}{2} \frac{\left(\pi ab\right)}{\left(4\pi L^{2} \right)}.
\end{equation} 
To minimize this ratio, Foucault pendula are generally made with
\textit{L} very large compared to the axes \textit{a} and \textit{b},
since it is comparatively easy to keep \textit{b} small while making
\textit{L} large.

Besides causing precession, ellipsoidal motion changes the
central oscillation frequency of the pendulum from $\omega _{0} $ to
\begin{equation} \label{GrindEQ__5_} 
\omega =\omega _{0} \left(1-\frac{1}{16} \frac{a^{2} +b^{2} }{L^{2} } \right).  
\end{equation} 
The fractional change in frequency due to a finite semi-minor axis
\textit{b }turns out to be of order $10^{-7}$, and therefore
negligible for the present discussion.

Every Foucault pendulum, no matter how carefully constructed to avoid
asymmetries in its suspension, and no matter how carefully
``launched'' to make ellipsoidal area \textit{A} as small as
reasonably achievable will, over time, acquire an intrinsic precession
$\Omega$ that can easily grow to overwhelm the Coriolis-force induced
Foucault precession $\Omega _{F} $.  Near $40^o$ latitude, for a
pendulum of length \textit{L} = 3.0 meters and semi-major axis
\textit{a} = 16. cm, the Foucault rate is equaled with a semi-minor
axis of \textit{b} = 3.9 mm.  The unwanted intrinsic precession may
add or subtract from the Foucault precession; when it is subtractive,
then the pendulum stops all precession when the corresponding value of
\textit{b} is reached.  In our experience, it is easy to launch such a
pendulum with semi-minor axis well under one millimeter, but under
free oscillation, on the time scale of only two to three minutes,
\textit{b} grows enough such that $\Omega$ dominates $\Omega_F$.

When the pendulum at an extremum, at $x=\pm a$ in Fig.~\ref{fig:1},
its motion is entirely transverse, with momentum $m\dot{y}$ as large
as it gets.  Preferential damping of this component of the motion will
reduce the unwanted ellipsoidal excursions.  In previous work, this
has been tried using a so-called Charron's ring around the suspension
wire near the top~\cite{cha1, cha2}, or letting a part of the pendulum
bob scrape an annular disk~\cite{cra1, cra2} at \textit{r = a}, or
using eddy current damping between a permanent magnet in the bob and a
non-ferrous metal annular disk~\cite{mas} located near \textit{x = a}.
We adopt the touch-free eddy-current damping method in the design we
present later in this paper.  Even in principle, none of these methods
will stop ellipsoidal motion completely, so an additional method is
needed to cope with any remaining intrinsic precession.

Every pendulum suffers dissipative losses of energy, mainly due to air
friction.  Long, very massive, museum-type pendula can simply be
relaunched once every day or so, but a small pendulum has a free
exponential decay time of order one half hour, so a ``driving''
mechanism is needed to restore the lost energy.  Various mechanisms
have been reported for this task~\cite{cra1,cra2,pri,mas}.  Like some
others, we will use magnetic induction to sense the passage near the
origin of a permanent magnet embedded in the pendulum bob.  A
carefully-timed electromagnetic impulse then imparts lost momentum to
the bob.  We will introduce a quantitative method for using a magnetic
push not only to compensate for dissipative losses, but also to
compensate for ellipsoidal precession.

The perturbations that lead to ellipsoidal motion are, in our
experience, in approximate decreasing order of severity, (1) internal
stresses or other imperfections in the fiber supporting the pendulum
bob, (2) less than perfectly symmetric suspension of the fiber at its
upper end, (3) nearby iron objects that result in asymmetric force on
the drive magnet in the bob, (4) a driving coil that is not
sufficiently level and centered under the pendulum.  All but the first
of these were straightforward to reduce to insignificance, but the
first was persistent.  This led to the necessity of finding a method
to evade the problem of ellipsoidal motion rather than to remove
it. There is no simple force law that leads to the intrinsic
precession $\Omega$, but rather it is an inescapable feature of the
spherical pendulum.  Nevertheless, we can apply a separate
perturbative force to counteract, i.e. nullify, the intrinsic
precession.  We discuss this in the next section.

\section{Method to Nullify Intrinsic Precession}

As shown in Fig.~\ref{fig:1}, the driving force that pushes the
pendulum away from the origin can be thought of as consisting of
components parallel and perpendicular to the major axis.  The parallel
component $F_{\parallel } $ is the larger one, and it is adjusted to
overcome the dissipative forces such as air resistance.  The
perpendicular component of the pushing force $F_{\bot}$ counteracts
the intrinsic precession, and we will now show that it is possible to
select distance \textit{d} at which the impulsive drive force is
applied to stop the precession $\Omega$.  The crucial result will be
that this distance is independent of semi-minor axis
\textit{b}. Starting from $x=-a$, in one half cycle of duration
${T\mathord{\left/ {\vphantom {T 2}}
\right. \kern-\nulldelimiterspace} 2} $, intrinsic precession advances
the ellipse in angle by ${\Omega T\mathord{\left/ {\vphantom {\Omega T
2}} \right. \kern-\nulldelimiterspace} 2} $. The pendulum arrives at
$x\simeq +a$ and
\begin{equation} \label{GrindEQ__6_} 
y\equiv \Delta y=\frac{1}{2} a\, \Omega T,  
\end{equation} 
where $\Delta y$ is the transverse displacement at the apex of the
ellipse.  However, we arrange to apply an impulsive momentum change at
$x=d$ that is calibrated to move the pendulum bob by a distance
$-\Delta y$ as it traverses the remaining longitudinal distance $a-d$.
If this is done, the bob will arrive at location $x=a$, $y=0$, as
desired.  The impulse and momentum change are related by
\begin{equation} \label{GrindEQ__7_} 
m\, \Delta v_{y} =F_{\bot } \Delta t,  
\end{equation} 
where $\Delta t$ is the duration of the applied force and \textit{m}
is the mass of the bob.  It turns out that $\Delta t$ is a few
milliseconds, compared to hundreds of milliseconds for the duration of
the swing, so it is appropriate to use the impulse formulation of
Newton's second law.  We treat this perturbation in the \textit{y}
direction as though the bob were free of other forces, which is
reasonable since $\Delta y$ is of order 10 microns compared to
\textit{a }of order 10 centimeters.  The horizontal component of the
motion is simply
\begin{equation} \label{GrindEQ__8_} 
x(t)=a\sin \omega _{0} t,  
\end{equation} 
where $\omega _{0} $ is the angular frequency characterizing the
pendular oscillations.  To good accuracy, the time $t_d$ between
passage closest to the origin and the instant the pendulum reaches
\textit{x = d} is thus
\begin{equation} \label{GrindEQ__9_} 
t_{d} =\frac{1}{\omega _{0} } \sin ^{-1} \frac{d}{a} .  
\end{equation} 
In one quarter of the full period \textit{T} the pendulum reaches its
apex.  Therefore, one way to write a relation between the distance
$\Delta y$ that the pendulum advances and the transverse velocity
$\Delta v_{y} $ that must be imparted by the driver is
\begin{align} 
\label{GrindEQ__10_} 
\Delta y=\Delta v_{y} \left(\frac{1}{4} T-t_{d} \right) & = 
\frac{\Delta v_{y} }{\omega _{0} } \left(\frac{\pi }{2} -\sin ^{-1} \frac{d}{a}
\right) \nonumber \\
&=\frac{\Delta v_{y} }{\omega _{0} } \cos ^{-1} \frac{d}{a}.
\end{align} 

From the geometry of the situation shown in Fig.~\ref{fig:1} we see
that that the driving force components are related by
\begin{equation} \label{GrindEQ__11_} 
\tan \theta =\frac{y}{d} =\frac{F_{\bot } }{F_{\parallel } } ,  
\end{equation} 
and, using the formula for an ellipse 
\begin{equation} \label{GrindEQ__12_} 
\left(x/a\right)^{2} +\left(y/b\right)^{2} =1,  
\end{equation} 
we have
\begin{equation} \label{GrindEQ__13_} 
F_{\bot } =F_{\parallel } \frac{b}{d} \sqrt{1-\left(\frac{d}{a} \right)^{2} } .  
\end{equation} 

Combining Eqs \eqref{GrindEQ__10_}, \eqref{GrindEQ__7_},
\eqref{GrindEQ__13_}, \eqref{GrindEQ__6_}, and \eqref{GrindEQ__2_} we
arrive at
\begin{align} \label{GrindEQ__14_} 
\Delta y &= \frac{1}{2} a\left(\frac{3}{8} \omega _{0} \frac{ab}{L^{2} }
\right)\left(\frac{2\pi }{\omega _{0} } \right) \nonumber \\
& = \frac{F_{\parallel }
\Delta t}{m\omega _{0} } \frac{b}{d} \sqrt{1-\left(\frac{d}{a}
\right)^{2} } \cos ^{-1} \frac{d}{a} .
\end{align} 
One sees in Eq \eqref{GrindEQ__14_} the crucial cancellation of the
factor \textit{b} that occurs between the intrinsic precession rate
and in the perpendicular component of the force.  This leads to the
result that follows being independent of the transverse size of the
ellipse.  This, in turn, means the result is insensitive to exactly
what non-central forces may act on the pendulum to cause non-vanishing
ellipsoidal motion.  An approximation is being made that $F_{\parallel}$ 
does not depend on \textit{b}; this will be justified later.

It is now convenient to introduce some dimensionless scaling
parameters.  Let
\begin{equation} \label{GrindEQ__15_} 
Q\equiv \frac{\pi }{2} \frac{ma\omega _{0} }{F_{\parallel } \Delta t} ,  
\end{equation} 
which is the ratio of the momentum of the pendulum as it crosses the
origin to the momentum kick it receives on each half oscillation.  The
denominator is the momentum the driver must supply to compensate for
the dissipative losses in order to maintain the full amplitude of the
swing.  The factor of $\pi /2$ stems from letting \textit{Q} represent
the conventional ``quality factor'' of an oscillator, specifically
that
\begin{equation} \label{GrindEQ__16_} 
Q=2\pi \frac{{\rm Total\; Energy}}{{\rm Energy\; Loss\; per\; Cycle}} .  
\end{equation} 
We expect this ratio to be quite large, on the order of 1000.  Then let
\begin{equation} \label{GrindEQ__17_} 
\alpha =\frac{a}{L}  
\end{equation} 
be the scaled amplitude of the pendulum, that is, the amplitude
divided by the length; this parameter is of order 0.1 for a practical
pendulum.  Next, let
\begin{equation} \label{GrindEQ__18_} 
\delta =\frac{d}{a}  
\end{equation} 
be the scaled distance from the origin at which the driving force is
applied, written as a fraction of the amplitude; this value can range
from near zero to unity.  With these dimensionless variables, Eq
\eqref{GrindEQ__14_} can be written
\begin{equation} \label{GrindEQ__19_} 
\boxed{\frac{3}{4} Q\alpha ^{2} =\frac{\sqrt{1-\delta ^{2} } }{\delta} 
\cos ^{-1} \delta }
\end{equation} 

\begin{figure}[ht]
\begin{center}
\includegraphics[width=0.48\textwidth]{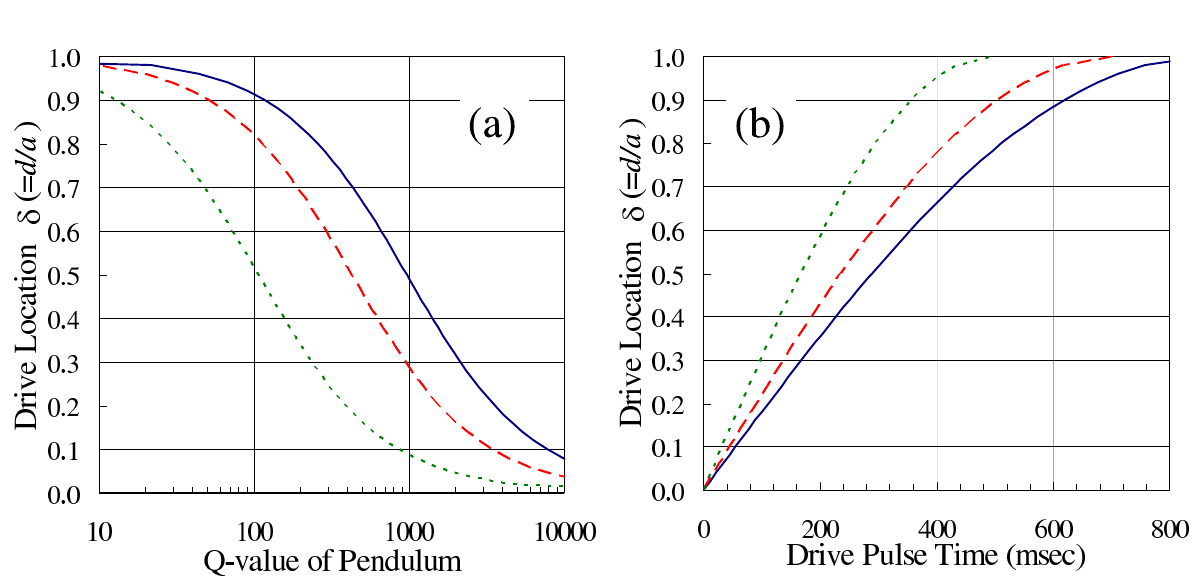}
\caption{
\label{fig:3}
(a) For three pendulum lengths (\textit{L}) with
amplitude of 0.15 meter, the relationship of the scaled driving
distance ($\delta=d/a$) versus the quality factor \textit{Q} for the
oscillation.  Curves are for an \textit{L} = 3.0 meter (solid blue),
2.0 meter (dashed red) and 1.0 meter pendulum (dotted green). (b) For
the same pendulum lengths as in (a), the distance versus time
relationship for the driving pulse.  }
\end{center}
\end{figure}

Equation \eqref{GrindEQ__19_} is the main result of this model.  It relates the scaled
distance $\delta$ at which the impulsive driving force must be applied
on each half oscillation, to the physical parameters of the pendulum,
specifically the scaled amplitude $\alpha$ and the quality of the
oscillator\textit{ Q}.  When Eq \eqref{GrindEQ__19_} is satisfied, the
intrinsic precession is stopped or nullified, and the result is
independent of the transverse size of the ellipse.
Figure~\ref{fig:3}a is an illustration of this result for the cases of
three different lengths of pendulum, each with a maximum excursion of
0.15 meters, as a function of the oscillation parameter, \textit{Q}.
It is seen that a low-loss pendulum with a larger value of \textit{Q}
will have to be pushed when it is closer to the origin, while a
pendulum with more dissipative losses, and therefore lower \textit{Q},
will have be driven further away from the origin.  A high-quality,
low-loss pendulum needs little energy input.  Therefore, the
transverse kick, which scales in magnitude with the longitudinal kick
at a given $\delta$, must be applied early in the ellipsoidal swing,
when the force has a larger transverse component.  Part (b) of
Fig.~\ref{fig:3} illustrates the relationship between the location of
the impulse, $\delta$, and the time, $t_d$, at which it is applied, as
per Eq \eqref{GrindEQ__9_}.

%
%

Taken together, the two parts of the Fig.~\ref{fig:3} enable design of
a Foucault pendulum not plagued by intrinsic precession.  For example,
a 3.0 meter long pendulum with a 0.15 meter amplitude ($\alpha
=0.050$) and a quality factor of $Q=1000$ must receive its impulsive
drive at $\delta =.49$, which corresponds to time $t_d = 284$ msec and
a drive location of \textit{d }= 7.4 cm from the origin.  On the other
hand, a 1.0 meter long pendulum with the same $\alpha$ and \textit{Q}
must receive its drive pulse at $\delta =.088$, at a time of 28 msec
and 1.3 cm from the origin.

As introduced in Eq \eqref{GrindEQ__15_}, \textit{Q} is proportional
to the ratio of the momentum of free oscillation at the origin to the
damping/driving momentum that accrues with each half period of
oscillation.  In fact, it is quite easy to measure this parameter for
an actual pendulum by measuring the period, \textit{T}, and the
exponential decay time, $\tau$. The motion of the pendulum in the
presence of velocity-dependent losses such as air friction is given,
to a good approximation at low speeds, by the damped harmonic
oscillator formula
\begin{equation} \label{GrindEQ__20_} 
x(t)=ae^{-t/\tau } \sin \omega _{0} t.  
\end{equation} 
The momentum loss between times \textit{t} = 0 and \textit{t = T}/2 is
\begin{equation} \label{GrindEQ__21_} 
m\dot{x}(0)-m\dot{x}(T/2)=ma\omega _{0} (1-e^{-T/2\tau } )\cong
ma\omega _{0} \frac{T}{2\tau}.
\end{equation} 
This leads to 
\begin{equation} \label{GrindEQ__22_} 
Q=\pi \frac{\tau }{T} ,  
\end{equation} 
from which \textit{Q }can be determined experimentally.  For the
actual three meter pendulum discussed below, we found the decay time
to be about 30 minutes and hence \textit{Q} to be roughly 1600.  Eq
\eqref{GrindEQ__22_} can be used with Eq \eqref{GrindEQ__19_} in order
to express Eq \eqref{GrindEQ__19_} in dimensioned variables as
\begin{equation} \label{GrindEQ__23_} 
\frac{3}{8} \sqrt{\frac{g}{L} } \left(\frac{a}{L} \right)^{2} \tau
=\frac{\sqrt{a^{2} -d^{2} } }{d} \cos ^{-1} \left({d\mathord{\left/
{\vphantom {d a}} \right. \kern-\nulldelimiterspace} a} \right).
\end{equation} 

One important approximation that was made needs to be examined.  In
going from Eq \eqref{GrindEQ__14_} to Eq \eqref{GrindEQ__19_}, the
cancellation of the semi-minor ellipse parameter \textit{b} was
crucial to showing that the result is independent of inevitable
changes in the size of the pendulum's ellipsoidal motion.  In fact,
$F_{\parallel } $ does depend very slightly on\textit{ b} in the
design we will discuss in the following sections.  The pushing agent
is a magnetic coil that produces a pulsed dipole field.  It acts
against a permanent magnetic dipole inside the pendulum bob.  Though
we operate in the near field of the coil, we presume that the magnetic
field has a distance dependence of the dipole form $B(r)=B_{0} (r_{0}
/r)^{3} $, where $B_{0} $ and $r_{0} $ are suitable scales.  The
dipole-dipole repulsion that drives the pendulum goes as the gradient
of this field, so the interaction has the form
\begin{equation} \label{GrindEQ__24_} 
\vec{F}=\vec{F}_{0} (r_{0} /r)^{4} ,  
\end{equation} 
where the components of $\vec{F}$ are what we earlier called 
$F_{\bot} $ and $F_{\parallel } $.  It is easy to expand this expression to
show that
\begin{equation} \label{GrindEQ__25_} 
F_{\parallel } =F_{0} \left(\frac{r_{0} }{d} \right)^{4}
\left\{1-\frac{5}{2} \left(\frac{b}{d} \right)^{2} \left(1-\delta ^{2}
\right)\right\}.
\end{equation} 
As seen in the second term in the curly brackets, there is a quadratic
dependence on the semi-minor axis \textit{b} that we ignored earlier.
This term is very small: for typical values of $\delta =d/a\simeq 6.0
cm/16.0 cm=0.375$, and $b/d\simeq 0.5 cm/6.0 cm = 0.083$, the second
term is 0.015, which is much less than 1.0.  Hence this approximation
was justified.

\section{Experimental Setup}

\begin{figure*}[ht]
\begin{center}
\includegraphics[scale=0.95]{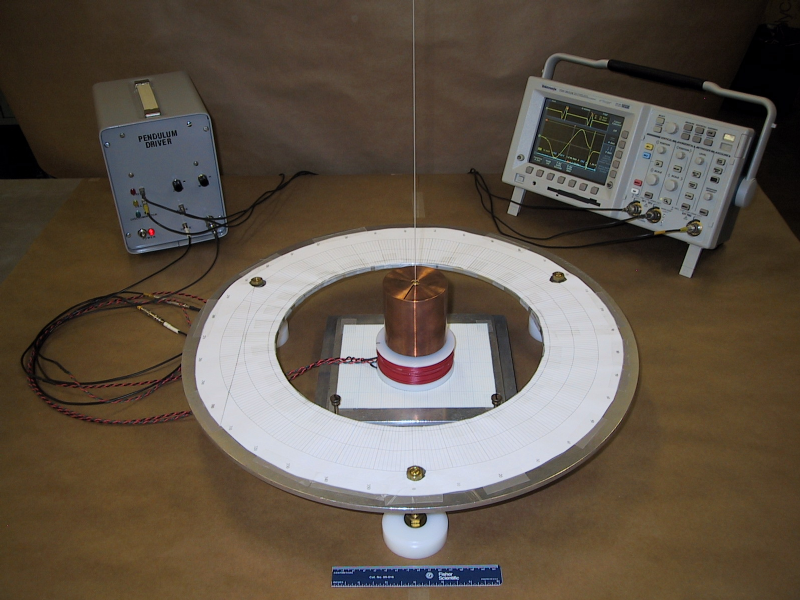}
\caption{
\label{fig:4}
Image of the copper pendulum bob suspended above the
sensing and driving coils (red wire on a white bobbin).  The aluminum
damping ring on brass legs is seen.  The white band carries angular
position markings.  The driver box and an oscilloscope are visible in
the background.}
\end{center}
\end{figure*}

To verify the mathematical model discussed above, a pendulum of about
three meter length was built.  The mechanics and drive electronics
will be described here.  An image of the lower end of the setup is
shown in Fig.~\ref{fig:4}.

%
%

The mass of the pendulum was in the form of a lathe-turned copper
cylinder, with a diameter of 2.75'' (6.99 cm) and height 3.00'' (7.62
cm).  A 3/8'' (.953 cm) axial hole was drilled and tapped to accept a
small threaded set screw with a 0.040'' (0.10 cm) hole in its center.
The small hole was for the fiber suspending the pendulum, and the
threading was used to adjust the distance between the center of mass
and the suspension point.  In practice, the fiber was held flush with
the top surface of the bob.  The bottom of the cylindrical bob was
chamfered by 0.10'' (0.25 cm) to allow closer approach to the damping
ring.  To hold the permanent magnet, a recess was milled into the
bottom of the mass, of diameter 3/4'' (1.90 cm) and depth 1/4'' (0.635
cm), which were the dimensions of the magnet.  The total mass of the
pendulum bob, apart from the magnet, was 2.47 kg.  This is a small
mass compared to what has generally been reported for Foucault
pendula.

The fiber suspending the mass was an ordinary and inexpensive polymer
material of the type used for yard trimmers.  The diameter was 0.040''
(0.10 cm).  The cross section of the fiber was circular and the
diameter was uniform to better than 1\%.  The polymeric material was
thought to be advantageous due to its lack of crystalline internal
structure.  It was found that the polymer material was very flexible,
yet strong, and did not suffer the bending fatigue and eventual
failure of some metal fibers we tried. In practice, we found that the
growth of elliptical motion of the pendulum was dominated by an
unobservable non-uniformity of this fiber, but it was less severe than
with various metal fibers. Tests wherein the fiber was rotated without
changing any other aspect of the pendulum showed this to be the case.
Thermal expansion and contraction of the fiber was large enough to
necessitate occasional shimming of the pendulum's length at the level
of about one millimeter.

The upper end of the fiber passed through a close-fitting drilled hole
in an aluminum plate.  The hole had a sharp rim, and no special steps
were taken to soften the bend of the fiber as it exited the hole.  On
the upper side of the plate the fiber was clamped in a way that thin
shims could be inserted to fine tune the length.  The plate was
leveled and clamped to rigid brackets on the ceiling of the
laboratory.  The upper support of the fiber used in this setup, though
carefully arranged, was certainly less exacting than what has been
found to be necessary for other Foucault pendula reported in the
literature.  We view this as an advantage of our design.

The permanent magnet placed inside the pendulum bob was a
neodymium-iron-boron disk magnet, placed with its dipole axis
vertical.  The axial field strength was 2.1 kGauss on contact, but it
varied by $\pm$10\% from one edge of the disk to the other.
Similarly, the radial field was 1.0 kG at the edge of the disk, but
varied also by $\pm$10\% from one side to the other.  Surprisingly
perhaps, these variations did not affect the performance of the
pendulum.

One purpose of the magnet was to provide eddy-current damping when the
bob passed over an aluminum ring near the extrema of its motion.  This
``damping ring'' had an inner diameter of 11 7/8'' (30 cm), and was
1/4'' (0.64 cm) thick.  Three legs made of threaded brass rod were
used to level the ring to a precision of less than half a millimeter
of variation around the perimeter.  The amplitude of the pendulum was
adjusted such that the magnet passed over the inner edge of the ring
with a clearance of three to four millimeters.  This maximized damping
of the unwanted transverse precessional speed. Since the eddy current
damping force is velocity dependent, it can never entirely stop this
motion, but our experience was that it limited the ellipsoidal motion
to a semi-minor radius of less than half a centimeter.  As a side
note, L\'eon Foucault first identified the eddy-current phenomenon in
1851, thus we use two phenomena in this investigation that are
attributed to him.

Two concentric coils under the pendulum sensed and controlled its
motion.  Both coils were wound from 22 AWG copper wire on polyethylene
bobbins, and both had 240 turns.  The inner coil was designated as the
``drive'' coil used for pulsing the pendulum on each half cycle.  It
had a mean radius of 0.80'' (2.0 cm) and a total length of 100' (30 m)
of wire.  The outer coil was the ``sense'' coil for detecting the
approach of the pendulum.  It had a mean radius of 1.5'' (3.8 cm) and
a total length of 187' (57 m) of wire.  The supply leads to both coils
were twisted pair copper wire near the coils, and RG-174 coaxial cable
at distances where the steel ground braid of the latter no longer
perturbed the motion.

%
%
\begin{figure}[ht]
\begin{center}
\includegraphics[width=0.40\textwidth]{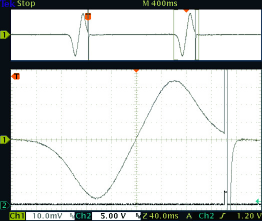}
\caption{
\label{fig:5}
Induced voltage at the input stage of the driver circuit, due to the
magnet in the pendulum approaching, passing over, and then receding
from the sense coil. Upper trace \eqref{GrindEQ__1_} shows a complete
cycle of two passages.  Lower zoomed trace shows how on the trailing
end of the sense pulse \eqref{GrindEQ__1_} the drive pulse
\eqref{GrindEQ__2_} causes a large bipolar induced pulse in the sense
coil. Screen capture was from a TDS 3032B oscilloscope.  }
\end{center}
\end{figure}

The concentric sense and drive coils were supported by a small
aluminum platform that sat on the main table, supported by three brass
screws with knurled heads.  These were used to level the coils:
imperfect leveling caused the induced signal from the pendulum,
described below, to be asymmetric.  The clearance between the bottom
of the pendulum and the top of the coils was 6$\pm$1 mm.

%
%
\begin{figure*}[ht]
\begin{center}
\includegraphics[scale=0.75]{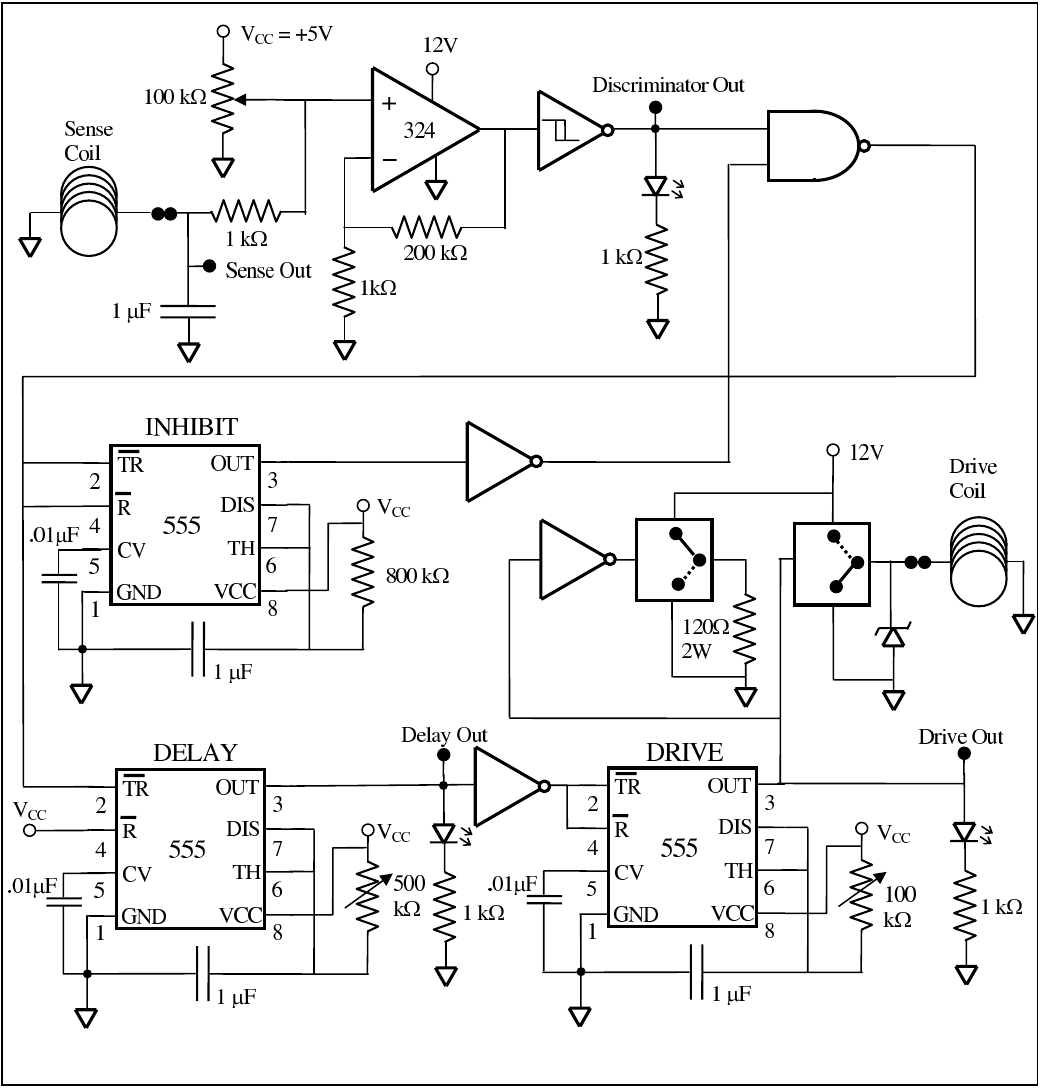}
\caption{
\label{fig:6}
Circuit diagram of the pendulum driver.  The op-amp (LM324) and
solid-state relays (Crydom CMX60D10) are supplied by +12V, while all
other components used $V_{CC}= +5$V.  Unlabeled circuit elements use
standard TTL chips.  The ``Out'' connections are for monitoring the
circuit on an oscilloscope.}
\end{center}
\end{figure*}

The sense coil acted to detect the induced EMF of the swinging
pendulum via Faraday induction.  The typical waveform is shown in
Fig.~\ref{fig:5}, showing that the approaching coil induced a negative
voltage that peaked at about $V_- = -34$ mV, as viewed across $1
M\Omega$ input impedance on an oscilloscope.  When the pendulum was
centered over the coil, the voltage crossed upwards through zero,
followed by a positive excursion of inverted shape and the same
magnitude $V_+$ as the negative part.  The peak magnitudes of the
waveform, V$_\pm$, depended on the speed of the pendulum (and hence
the amplitude of the swing), as well as the distance between the
pendulum magnet and the coil.  Monitoring V$_\pm$ over time was a
sensitive way to detect small changes in the pendulum's amplitude
and/or length.

The pendulum driver circuit developed for this study is shown in
Fig.~\ref{fig:6}, and operated as follows.  The initial signal from
the coil was filtered with a 1 $\mu$F capacitor that reduced RF noise
to less than one millivolt.  An op-amp circuit of type LM324N
level-shifted and amplified the signal by a factor of 200. This signal
was the input to a SN7414N Schmitt trigger that switched at 0.9 and
1.7 Volts.  In effect, this was a discriminator with a TTL output
``trigger'' pulse when the induced voltage went more negative than
$-20$ mV; the trigger stayed high until the trailing end of the
sense-coil signal shown in Fig.~\ref{fig:5} fell.  A NAND gate output
was used to trigger two 555 timers called DELAY and INHIBIT; the
second input of the NAND was an inhibit signal that ensured that the
DELAY timer was not restarted by the off-scale pulse induced by
activation of the drive coil.  The DELAY timer set the interval
between the approach of the pendulum and the drive pulse in the second
coil; it was adjustable in the range from near zero to several hundred
milliseconds.  The INHIBIT timer duration was set to be longer than
the maximum possible delay plus drive times.  The output of the
INHIBIT timer was fed via an inverter back to the NAND, so the drive
pulse could not re-fire the delay timer.  The output of the DELAY
timer was passed via an inverter to the input of the third 555 timer
called DRIVE.  This timer was adjustable over a range of several tens
of microseconds, and was used to set the duration of the current
through the drive coil.  The output of the DRIVE timer was used to
control two high-current MOSFET relays that switched current to and
from the drive coil.  The circuit was designed to maintain a
substantial current draw from the commercial 12V switching power
supply at all times, hence the dual opposite-acting relays.  A Zener
diode across the drive coil prevented back-emf from damaging the relay
and power supply during switching.  The very brief drive current was
estimated to be 7.5 Amps.  The circuit was constructed using the
wire-wrap method and housed with its power supply in a small box.
External connections were made using LEMO-style coaxial connectors, as
seen in Fig.~\ref{fig:4}.

\section{Results}

The pendulum described in the preceding section was used to determine
the accuracy to the model that lead to Eq \eqref{GrindEQ__19_}.  The
prediction that the Foucault precession rate should be unaffected by
the transverse size of the elliptical motion is to be confirmed.

The measured pendulum parameters were $T=3.44\pm 0.01$sec, and
$a=16.2\pm 0.2$cm.  The free decay period was determined by making an
exponential fit to amplitude-vs.-time data, leading to $\tau =29.7\pm
0.5$min.  From these values we computed $L=2.938\pm 0.009$m, $\alpha
=0.0551\pm 0.0007$, and $Q=1628\pm 28$.  Solving Eq
\eqref{GrindEQ__19_} numerically leads from these values to $\delta
=0.319\pm 0.006$.  This value in turn can be converted using Eq
\eqref{GrindEQ__9_} to an expected drive time $t_{d} =178\pm 3$ms.  On
the other hand, the actual value of $t_{d}$ was adjusted such that a
steady Foucault precession, as shown below, was observed.  The direct
measurement of this time using an oscilloscope gave $t_{d} =180\pm
2$ms.  Thus, the model expectation as given in Eq \eqref{GrindEQ__19_}
is found to be in excellent agreement with experiment with about 1\%
precision.

\begin{figure}[ht]
\begin{center}
\includegraphics[width=0.50\textwidth]{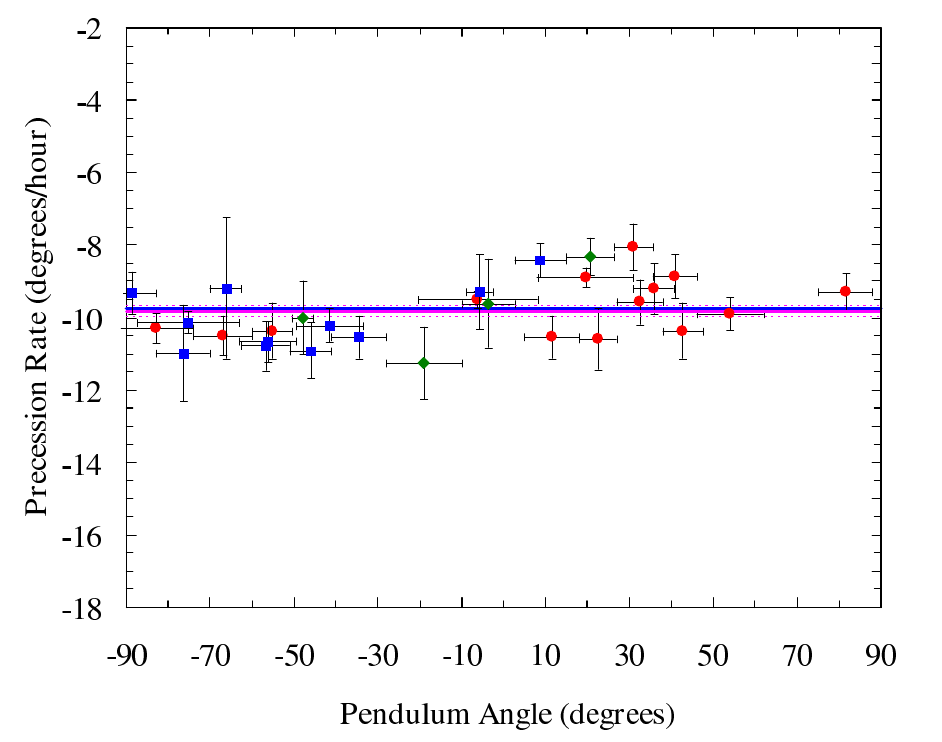}
\caption{
\label{fig:7}
Precession rate as a function of azimuthal
angle of the pendulum.  Red circles are for counterclockwise, while
blue squares are for clockwise ellipsoidal perturbation.  Green
diamond points were for measurements with no discernible elliptical
motion. The expected rate is shown as the thick blue line. The
measured average precession rate (solid) and its uncertainty band
(dotted) are shown as red lines.
}
\end{center}
\end{figure}

Figure~\ref{fig:7} shows the result of a set of measurements of the
precession rate spanning several days.  The abscissa angles were
measured from an origin arbitrarily picked to point north, with
positive azimuthal angles to the counterclockwise side, as seen from
above.  Red circular data points are for counterclockwise and blue
squares are for clockwise ellipsoidal excursions in the pendulum's
motion.  Green diamond points are for readings with no measurable
sense, i.e. for $b\simeq 0$.  At the latitude of $\theta _{latitude}
=40^{{\rm o}} 26'26"$ (Pittsburgh, Pennsylvania), the expected
Foucault precession rate in one sidereal day is
\begin{equation} \label{GrindEQ__26_} 
\Omega _{F} =-\frac{2\pi }{23.935{\rm \; hours}} \sin \theta
=-0.1703/{\rm hr}\to -{\rm 9.757}^{{\rm o}} /{\rm hr},
\end{equation} 
and this is shown as the blue bar.  The vertical error bars on the
points represent the estimated random measurement errors, and these
were dominated by a $\pm 1/2$ degree precision of the angle
measurements.  The horizontal error bars represent the span of angles
over which the rate was measured, with a typical span being ten
degrees.  The scatter of the data points clearly clusters around the
expected rate, with no trend in angle or sense of ellipsoidal motion.
The simple average of the measured points gives $\Omega _{F} =-9.85\pm
.15$, where the given uncertainty is the error on the mean, not the
standard deviation, and this is the red-dotted band shown in the
figure.  The weighted mean of the data gives $\Omega _{F} =-9.69\pm
.09$, which is in very good agreement with the simple average, and
both are in excellent agreement with the expectation of Eq
\eqref{GrindEQ__26_}.  We prefer the former uncertainty because the
wide scatter of the data points suggests that perhaps there are some
small systematic effects that are not captured using the weighted
mean's uncertainty.  Hence the former uncertainty is a better estimate
of the true uncertainty of the result.  The main candidate for a
poorly-controlled systematic effect is the temperature-related changes
in the length of the pendulum.  Even at the sub-millimeter level,
these affected the strength of the magnetic kick the pendulum
received, and hence affected the value of $\alpha$, which in turn
could affect control of the precession rate. We fine-tuned the length
of the pendulum with shims, as needed, to minimize this effect, but
the compensation was not perfect, and this probably led to some loss
of reproducibility.  Nevertheless, we have found the expected Foucault
precession rate with an angle-integrated precision of 1.5\%.

%
%

%
%
\begin{figure}[ht]
\begin{center}
\includegraphics[width=0.50\textwidth]{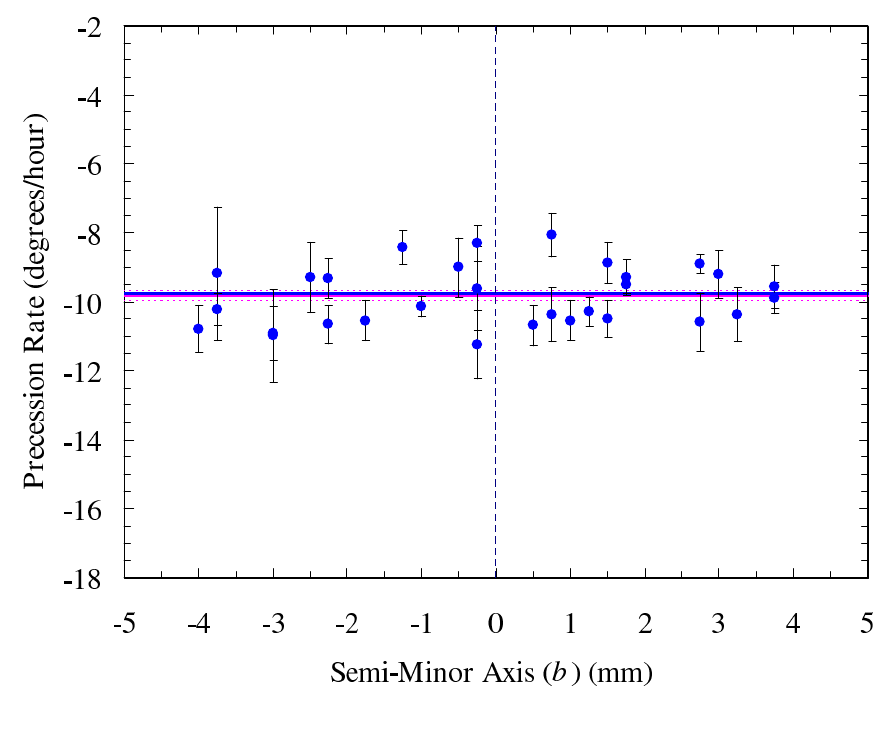}
\caption{
\label{fig:8}
Precession rate as a function of the size
of the semi-minor axis \textit{b }of elliptical motion.  Positive
values of \textit{b} are for counterclockwise ellipses, negative
values for clockwise ellipses.  The horizontal lines are the same as
in the previous figure.
}
\end{center}
\end{figure}

As discussed in connection with Eq \eqref{GrindEQ__14_}, if $t_d$ is
correctly chosen, the rate of precession in this pendulum should not
depend on the size of the semi-minor axis, \textit{b}.  We could not
control \textit{b} since it depended on the small asymmetries of the
fiber itself and any other perturbations of the pendulum's motion.  Of
course it was damped by the effect of eddy current braking against the
damping ring, but there was always some irreducible amount of
ellipsoidal motion to contend with.  We assign clockwise motion to
have negative \textit{b} values and counterclockwise motion to have
positive \textit{b} values.  Figure~\ref{fig:8} shows our measured
precession rates, the same data set as in the previous figure, as a
function of \textit{b}.  The size of \textit{b} was measured by
watching the pendulum pass over a ruler placed under the pendulum, and
the precision of doing this was no better than $\pm 1$mm. We plot the
points at the mean value of the ellipse size during the measurement
interval.  It is clear from the figure that there is no correlation
between the two variables, thus proving the claim that they are in
fact independent when the pendulum is properly arranged.  The Foucault
precession was clockwise in this northern-hemisphere setup,
\textit{i.e.} toward decreasing angles.  Note that even when the
pendulum was moving in a counterclockwise ellipse (positive values of
\textit{b}) that would normally precess in a counterclockwise
direction, the action of the driver was such that the clockwise
Foucault precession rate was unaffected.  This again shows how our
method nullifies the unwanted intrinsic precession.\textbf{}

\begin{figure}[ht]
\begin{center}
\includegraphics[width=0.50\textwidth]{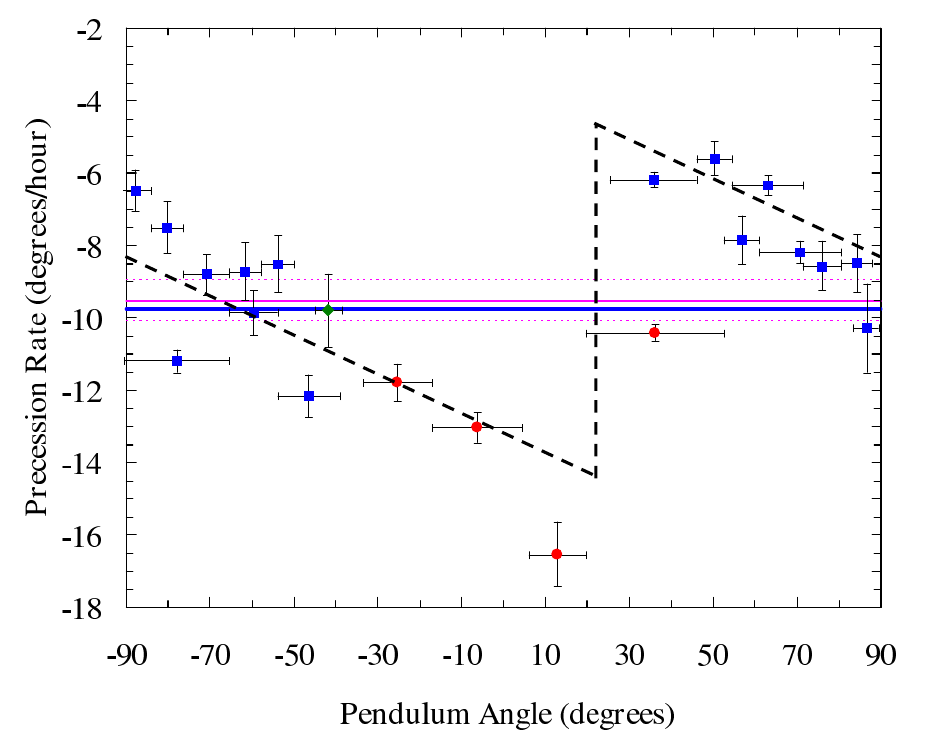}
\caption{
\label{fig:9}
Precession rate as a function of azimuthal angle of the
pendulum when the driving time $t_d$ is reduced by $\approx20\%$ from
the ideal value.  As in Fig.~\ref{fig:7}, red circles are for
counterclockwise, while blue squares are for clockwise ellipsoidal
motion.  Green diamond points were for measurements with no
discernible elliptical motion.  The expected rate is shown as the
thick blue line. The measured average precession rate (solid) and its
uncertainty band (dotted) are shown as red lines.  The heavy black
dashed line is a guide to the eye, as discussed in the text.  }
\end{center}
\end{figure}

As further evidence that our method succeeds in controlling the motion
of the pendulum, we intentionally reduced the value of the driving
time by $\approx20\%$, so that the transverse kick given by $F_{\bot
} $was larger than optimal.  This means that any ellipsoidal
precession was overcompensated by the drive system, resulting in
forced precession in the opposite sense to what free oscillation would
produce. Figure~\ref{fig:9} shows the precession rate as a function of
angle for $t_{d} =147\pm 1$ msec.  There is, in this new situation,
substantial systematic deviation from the expected rate as a function
of azimuthal angle.  The angle-integrated average rate is now $\Omega
_{F} =-9.53\pm .57$, which is still in agreement with the expected
Foucault rate, but now with a much wider error band, as shown.  One
also sees the propensity for clockwise ellipsoidal motion (blue
squares) to decrease the magnitude of the precession rate, while
counterclockwise motion (red circles) increases the magnitude of the
precession rate.  This is as expected in our mathematical model.
There are again some cases of irreproducibility of the data points at
a given angle.  We believe this is the result of occasional shimming
of the pendulum's length during a multi-day run; it leads to
uncontrolled small variations in the performance of the system.

%
%

The heavy black dashed line in Fig.~\ref{fig:9} is a guide to the eye
to illustrate what happens when the ellipsoidal precession is not
perfectly canceled.  There is a preferred axis near $-67^o$, where the
measured rate crosses the Foucault rate, because here the non-central
forces present in the system happen to vanish.  For more negative
azimuthal angles (which then wrap to the positive side of the diagram)
the precession rate is slower.  This is because the unwanted forces
act to cause clockwise ellipsoidal motion, but the drive system is
overcompensating and adding a component of counterclockwise
precession.  At about $+23^o$, which is $90^o$ away from the preferred
axis, there is a tipping point where the non-central forces act to
push the pendulum in the other direction, counterclockwise. But here
the overcompensating drive system makes the pendulum precess more
quickly clockwise.  We do not claim that the departure from the
Foucault rate is strictly linear, as shown, only that the trends are
consistent with our understanding of the physical process.

\begin{figure}[ht]
\begin{center}
\includegraphics[width=0.50\textwidth]{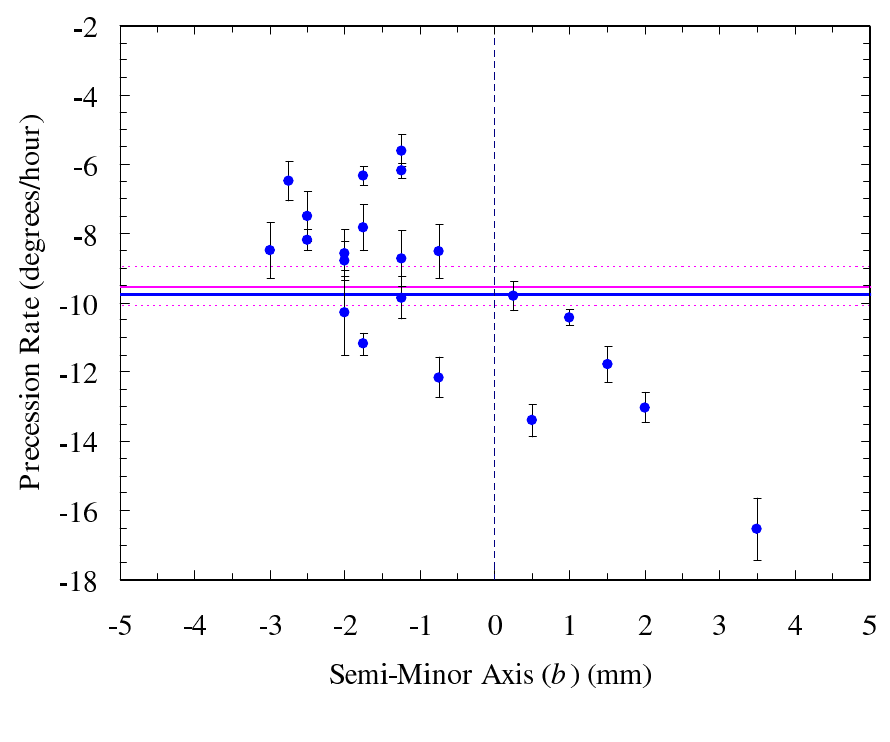}
\caption{
\label{fig:10}
Precession rate as a function of the size of the
semi-minor axis \textit{b }when the driving time $t_d$ is reduced by
$\approx20\%$ from the predicted value.  Positive values of
\textit{b} are for counterclockwise ellipses, negative values for
clockwise ellipses.  This figure is to be compared with
Fig.~\ref{fig:8}.  }
\end{center}
\end{figure}

Finally, in Fig.~\ref{fig:10} we show the effect of reducing the drive
time $t_d$ by $\approx 20\%$ on the relationship between precession
rate $\Omega$ and the semi-minor axis \textit{b}.  One sees clearly a
correlation between them, whereas in Fig.~\ref{fig:8} there was no
correlation.  For positive values of \textit{b} (counterclockwise
ellipses), the precession rate has increased magnitude since the
overcompensating drive system ``adds'' to the Foucault rate.  For
negative $b$ values (clockwise ellipses) the precession rate is
decreased in magnitude since the overcompensating drive system
``subtracts'' from the Foucault rate.  Thus, we have shown that when
deviating from the prediction of Eq \eqref{GrindEQ__19_} for the
correct drive distance, and therefore the correct drive time, the
pendulum shows marked departure from the constant Foucault precession
rate that is expected.

%
%

\section{Further Discussion and Conclusions}

In the previous work of Crane~\cite{cra2}, he recognized the
importance of nullifying the intrinsic precession that remains after
damping as well as possible.  However, he used a ``push-pull'' drive
system to mitigate problems of alignment of the coils with the
pendulum.  This led him to introduce a carefully-placed fixed
permanent magnet at the origin to provide the desired stopping of
intrinsic precession. No quantitative understanding of how to predict
the placement of this magnet was offered.  It seems to us that his
very delicate \textit{ad hoc} adjustment of this auxiliary magnet is
difficult, and, as we have shown, not necessary.  Our method is
simpler and more direct, in that it does not require this additional
magnet.  Alignment of the driving and sense coils was not found to be
a problem, and the results were not sensitive to the alignment at the
level of about a millimeter.  This was because the method we have
introduced is in a sense self-correcting: if the drive coil causes
some small amount of ellipsoidal motion, the very action of the method
prevents this motion from causing unwanted precession.  The work of
Mastner \textit{et al.}  ~\cite{mas} made note of the benefits of a
``push only'' driving force, but they did not offer the quantitative
explanation as to how and why it worked.  They built a traditional,
very long, very massive pendulum, taking great care to minimize
asymmetries.

In conclusion, we have demonstrated the workings of a ``short''
Foucault pendulum that was designed on a quantitative basis to avoid
the unwanted precession due to ellipsoidal motion.  The design formula
we have derived, Eq \eqref{GrindEQ__19_}, was shown to agree with
measured values in our setup.  We have shown that a driving mechanism
that pushes, not pulls, the pendulum is the key to canceling the
intrinsic precession for all values of the semi-minor axis of the
ellipse.  Our driver system used Faraday induction and magnetic
repulsion to control the pendulum, using a circuit based on simple
op-amp, logic, and timer chips.  Eddy current damping was used to
reduce the ellipse size, but active compensation did the rest.  The
design is immune to small non-central forces that are difficult to
control in a short pendulum.  We plan to further test this method on
even shorter pendula, since there is no lower limit at which the model
given here should apply.

\section{Acknowledgments}

We thank Mr.~Gary Wilkin for his expert help in the machine shop.  We
thank Mr.~Michael Vahey for construction of the pendulum driver, and
we thank Mr.~Chen Ling for help with exploratory initial trials in
construction of a Foucault pendulum.



\begin{thebibliography}{5}

\bibitem{fou} M. L. Foucault, ``D\'emonstration physique du mouvement de
rotation de la terre au moyen du pendule'' Comptes Rendus
Acad. Sci. {\bf 32}, 135-138 (1851).

\bibitem{ols1} M. G. Olsson, ``The precessing spherical pendulum,''
Am. J. Phys. {\bf 46}, 1118-1119 (1978).

\bibitem{ols2} M. G. Olsson, ``Spherical pendulum revisited,''
\noindent Am. J. Phys. {\bf 49}, 531-534 (1981).

\bibitem{pip} A. B. Pippard, ``The parametrically maintained Foucault
pendulum and its perturbations,'' Proc. R. Soc. Lond. {\bf A420}, 81-91 (1988).

\bibitem{syn} J. Synge and B. Griffith, {\it Principles of Mechanics}
(McGraw Hill, 1959), 3rd ed., pp 335-342.

\bibitem{cha1} M. Charron, ``Sur un perfectionnement du pendule de
Foucault et sur l'entretien des oscillations,'' Comptes Rendus
Acad. Sci. {\bf 192}, 208-210 (1931).  

\bibitem{cha2} See for example C.F. Moppert and W. J. Bonwick, ``The
New Foucault Pendulum at Monash University'',
Q. Jl. R. Astr. Soc. {\bf 21}, 108-118 (1980), and references therein.  Also
Haym Kruglak {\it et al}, ``A short Foucault pendulum for a hallway
exhibit'', Am. J. Phys. {\bf 46}, 438-440 (1978).

\bibitem{cra1} H. Richard Crane, 
``The Foucault Pendulum as a murder weapon and a physicist's delight'', 
Phys. Teach. 264-269 (May 1990).

\bibitem{cra2} H. Richard Crane, ``Foucault pendulum ``wall clock'''', 
Am. J. Phys. {\bf 63}, 33-39 (1995); 
``Short Foucault Pendulum: A Way to Eliminate the Precession due to Ellipticity'', 
Am. J. Phys. {\bf 49}, 1004-1006 (1981).

\bibitem{mas} G. Mastner {\it et al}, ``Foucault pendulum with eddy-current
damping of the elliptical motion'', Rev. Sci. Inst. {\bf 55}, 1533-1538 (1984). 

\bibitem{pri} Joseph Priest and Michael Pechan, ``The driving mechanism for a 
Foucault pendulum (revisited)'', Am. J. Phys. {\bf 76}, 188-188 (2008).

\end{thebibliography}
\end{document}